\documentclass[reprint,prd]{revtex4-1}
 \usepackage{axodraw}
 \usepackage{amssymb}
 \usepackage{graphicx}

 \newcommand{\jouref}[4]{{#1 }{\bf #2} (#3) #4}
 
 \newcommand{\hepph}[1]{{hep-ph/#1}}

\begin{document}

 \title{Tadpole cancellation in top-quark condensation}

 \author{Diganta Das}
 \affiliation{The Institute of Mathematical Sciences, CIT Campus, 
Taramani, Chennai 600 113, India}

 \author{Kosuke Odagiri}
 \affiliation{Electronics and Photonics Research Institute,
 National Institute of Advanced Industrial Science and Technology,
 Tsukuba Central 2,
 1--1--1 Umezono, Tsukuba, Ibaraki 305--8568, Japan}

 \date{August 2012} 

 \begin{abstract}
  We show that quadratic divergences in top-quark condensation are 
cancelled when the tadpoles cancel.
  This latter cancellation is naturally implemented as the cancellation 
among the top-quark, Goldstone and Higgs contributions.
  We also calculate the bosonic correction terms to Gribov's mass 
formula for the Higgs boson. These reduce the prediction for $M_H$ from 
$167$~GeV to $132$~GeV.
  The tadpole cancellation condition by itself is an independent 
condition on the mass of the Higgs boson which, in Gribov's U(1)$_Y$ 
scenario, yields $M_H\approx117$~GeV with large theoretical uncertainty.
  More generally, we are able to obtain all three masses, $M_W$, $m_t$ 
and $M_H$, in $100$~MeV to $10$~TeV energy range as a function of the 
cut-off scale and the gauge couplings only.
 \end{abstract}


 \maketitle

 \section{Introduction}

  As is well known, the Standard Model, with an elementary Higgs doublet 
field, suffers from the problem of quadratic divergences when radiative 
corrections, due to the loops of fermions, the top quark in particular, 
to the mass of the Higgs boson is calculated.
  This problem is often referred to as the hierarchy problem, as we 
require an artificial fine tuning between the induced radiative mass, 
which is of the order of the cut-off scale, e.g., 
$\mathcal{O}(M_\mathrm{Pl})$, and the counterterm to produce a mass of 
the order of the electroweak scale.

  This tempts us to speculate that the Higgs boson may in fact be 
composite.
  The simplest implementation of this idea is top-quark condensation 
\cite{nambu,bardeen,topcondensation}.
  It turns out, however, that this compositeness by itself is not 
sufficient to remove the problem of quadratic divergences, and fine 
tuning is still required in the simpler approaches to top-quark 
condensation.

  The problem of divergences can only be artificial, because the same 
loop corrections, applied to the mass of the Goldstone bosons, produce a 
similar quadratic divergence, whereas the Goldstone theorem guarantees 
that spontaneous symmetry breaking results in massless Goldstone bosons.
  This suggests that these quadratic divergences are an artifact, and 
will vanish if the condition of current conservation is implemented 
properly.

  For example, the approach of Chesterman, King and Ross 
\cite{bethe_salpeter} uses the vanishing of the mass of the Goldstone 
boson as a consistency check, and is able to obtain sensible values for 
the mass of the Higgs boson.

  A more direct investigation into this point regarding current 
conservation was made by Gribov in ref.~\cite{gribovewsb}, and his 
results are in quantitative agreement with ref.~\cite{bethe_salpeter}.
  In that paper, he implemented the symmetry condition, somewhat by 
force, by requiring that the Goldstone-boson self-energy vanishes in the 
soft limit.
  The mass of the Higgs boson is then obtained by subtracting off the 
Goldstone-boson self-energy, leading to the following 
Pagels--Stokar-type equation for the mass of the Higgs boson:
 \begin{equation}
  M_H^2=\frac{N_C}{2v^2\pi^2}\int_{m_t^2}^{\lambda^2}
  \frac{dq^2}{q^2}m_t^4(q^2).
  \label{eqn_gribov_higgs_mass}
 \end{equation}
  This gives the mass $M_H=167$~GeV for the Higgs boson using the 
top-quark mass $m_t=174$~GeV and $\alpha_S=0.11$ as input.
  Gribov's cut-off $\lambda$ is given by the U(1)$_Y$ Landau pole, 
$\sim10^{42}$~GeV, but the results are relatively insensitive to the 
value of the cut-off scale.
  This value of $M_H$ is not far from the recent LHC results 
\cite{cernpressrelease,atlashiggs,cmshiggs} which suggest 
$M_H\approx125$~GeV.
  It is worth noting here that eqn.~(\ref{eqn_gribov_higgs_mass}) is 
general and is independent of the physics that causes the cut-off.

  Our questions are the following. First, what may be the justification 
for this procedure, in particular the mechanism for the cancellation of 
the divergences? Second, whether there may be correction terms due to 
the loops of Goldstone and Higgs bosons in 
eqn.~(\ref{eqn_gribov_higgs_mass}).

  Concerning the first question, we shall show in this paper that the 
quadratic divergences are proportional to the tadpole and therefore 
vanish when the vacuum is stable.
  Although this statement must be independent of the basis, it is most 
easily verified by defining the Goldstone bosons as the divergence of 
the weak current so that derivative couplings do not arise.

  Our analysis is quite general in the sense that no new interactions 
are introduced up to the UV cut-off scale.
  This procedure has the well-known phenomenological disadvantage that 
the mass $M_W$ of the $W$ boson is predicted to be too low for fixed 
$m_t$ or, equivalently, $m_t$ is predicted to be too large for fixed 
$M_W$. This is partially alleviated by taking large $\lambda$ such as, 
as in Gribov's scenario, the U(1)$_Y$ Landau pole. Even this is not 
sufficient and, as remarked in ref.~\cite{gribovewsb}, more UV 
contribution is required, which may be, one would naturally guess, due 
to strong dynamics at near and above the cut-off.

  The methods of our analysis are analogous to those in 
ref.~\cite{odagirimagnetism} that are worked out in the context of 
anti-ferromagnetism.
  The second of the questions above is also addressed within this 
framework, and we find that the correction to 
eqn.~(\ref{eqn_gribov_higgs_mass}) is large and negative.

  As an explanation of our framework, we implement current conservation 
by requiring the vanishing of the Ward--Takahashi identities. This fixes 
the bosonic three-point and four-point couplings.
  In other words, the shape of the Higgs potential is fixed by current 
conservation, and this turns out to be of the same form as that of the 
Standard Model.
  We then implement the vacuum-stability condition, namely that the 
Higgs $\to X$ one-point function, or the tadpole, vanishes.

  Note that the cancellation of the tadpole, which is imposed as the 
cancellation among the top-quark, Goldstone-boson and Higgs-boson loops, 
is by itself an independent condition on the symmetry-breaking 
parameters. This yields
 \begin{equation}
  M_H^2\approx8m_t^2(\lambda^2),
  \label{eqn_tadpole_higgs_mass}
 \end{equation}
  which implies $M_H\approx117$~GeV in the U(1)$_Y$ scenario, if we only 
include the QCD part of the running.
  This is very close to the LHC value, but this remarkable success 
should not be taken too seriously, because this condition will be 
affected by the renormalization, due to the U(1)$_Y$ interaction, of the 
top-quark mass at near the Landau pole.

  It should be noted that this extra condition on the Higgs mass, in 
principle, completely fixes the symmetry-breaking parameters.
  That is, all three masses, $M_W$, $m_t$ and $M_H$, are fixed for given 
values of the cut-off scale and gauge couplings.
  This provides a natural solution to the hierarchy problem and, indeed, 
we obtain a large hierarchy that is comparable with the actual hierarchy 
between the UV and EWSB scales.

  The form of the UV theory does not affect our analysis at the present 
level of approximation, so long as the symmetry is spontaneously broken 
by a UV condensate which exists only at high energies.
  This requires a supercritically strong UV interaction, and a natural 
candidate is Gribov's U(1)$_Y$ scenario.
  We discuss the physics of this scenario.

 \section{Current conservation conditions}

  Tadpole cancellation in bosonic self-energy was, in part, discussed by 
Gribov himself in the context of chiral symmetry breaking in 
ref.~\cite{gribovlongshort}. However, he imposes tadpole cancellation 
there by introducing a boson--boson--quark--quark term (e.g., eqn.~(132) 
of ref.~\cite{gribovlongshort}) to cancel the anomalous term in the 
Ward--Takahashi identity associated with Compton scattering.
  One problem in doing so is that this will lead to the necessity of 
introducing an infinite number of $N$-boson--quark--quark couplings.

  In ref.~\cite{odagirimagnetism}, and in the context of 
anti-ferromagnetism, we proposed a more economical mechanism which also 
seems to us to be more natural.
  Here, only the three-point and four-point couplings among the Goldstone
and Higgs bosons are needed, and these couplings are 
fixed by symmetry conditions such as the Ward--Takahashi identities.

  We shall demonstrate that the couplings are consistent with the 
Standard-Model-like quartic Higgs potential.
  In the low-energy effective theory, the phenomenology is identical to 
that of the Standard Model, at least insofar as the couplings to the 
bosons and the top quark are concerned. As for the other fermionic 
Yukawa couplings, it is natural to assume that they are as given by the 
Standard Model though, strictly speaking, other possibilities cannot be 
ruled out.

  Let us start with the fermionic vertex function for the left-handed 
SU(2) current.
  As discussed in ref.~\cite{gribovewsb}, Ward--Takahashi identity is 
satisfied by the following modified vertex:
 \begin{equation}
  \includegraphics{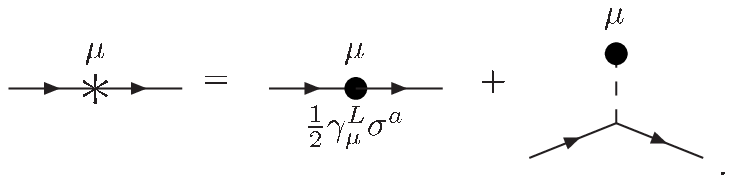}
  \label{eqn_modified_vertex}
 \end{equation}
  Here, the small blob stands for the unmodified vertex 
$\frac12\gamma_\mu^L\sigma^a$ and the two-point function ($f_{ab}p_\mu$ 
in Gribov's notation) which follows from that, and the asterisk stands 
for the vertex that is modified by the inclusion of the Goldstone-boson 
contribution.
  The dashed line stands for one of the three Goldstone bosons $G^0$, 
$G^\pm$.

  The Ward--Takahashi identity, applied to the modified vertex, fixes 
the $G$ coupling to fermions to be, for example,
 \begin{equation}
  g_{G^0}=i\gamma_5\sigma^3\frac{m}{2f},
  \label{eqn_goldstone_fermion_coupling}
 \end{equation}
  where $f$, the Goldstone-boson form factor, is defined by the 
two-point function of eqn.~(\ref{eqn_modified_vertex}). $f=v/2$, where 
$v$ is the usual `vacuum-expectation value of the Higgs field'.
  Note that the Feynman rules are, as usual, given by $-ig$.
  This will apply to all couplings that appear in the following.

\begin{figure}[ht]
  \includegraphics{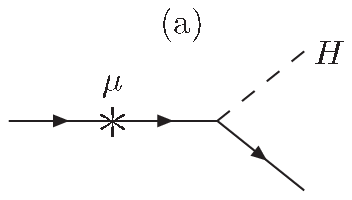}
  \includegraphics{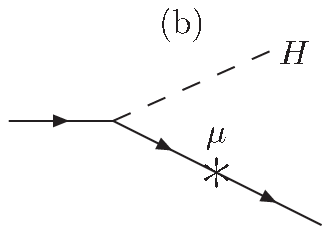}

  \includegraphics{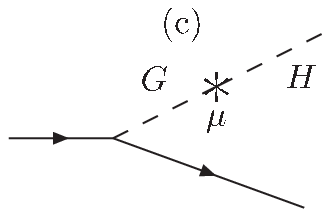}
  \caption{\label{fig_fermion_boson_current_amplitude}
  The three diagrams whose sum must satisfy the Ward--Takahashi 
identity.}
 \end{figure}

  The same exercise may be repeated for the bosonic vertex:
 \begin{equation}
  \includegraphics{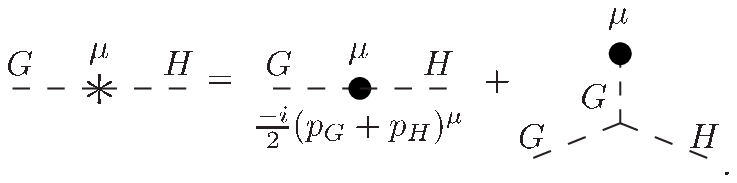}
  \label{eqn_modified_vertex_bosonic}
 \end{equation}
  The unmodified vertex is necessarily proportional to $p_G+p_H$ 
(momenta flows left to right, or more generally $G$ to $H$) in order to 
satisfy the Ward--Takahashi identity, and the normalization is fixed by 
the Ward--Takahashi identity applied to the amplitude shown in 
fig.~\ref{fig_fermion_boson_current_amplitude}.

  This fixes the $GGH$ coupling to be:
 \begin{equation}
  g_{GGH}=\frac{M_H^2}{2f},
 \end{equation}
  when the Yukawa coupling of the Higgs boson to fermions is given by
 \begin{equation}
  g_{H}=\frac{m}{2f}.
  \label{eqn_higgs_fermion_coupling}
 \end{equation}

 \begin{figure}[ht]
  \includegraphics{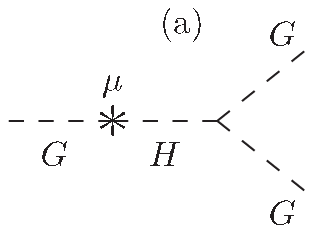}
  \includegraphics{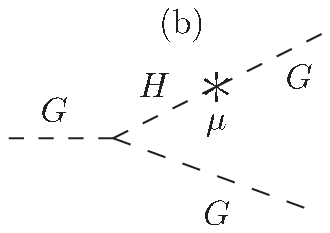}

  \includegraphics{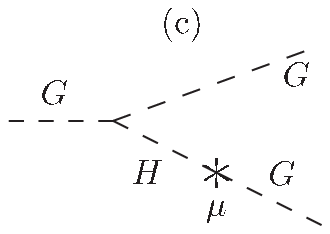}
  \includegraphics{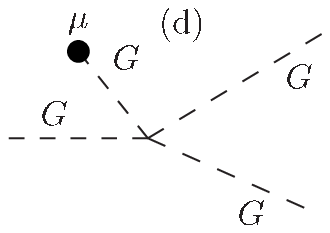}
  \caption{\label{fig_triple_boson_current_amplitude}
  Four diagrams whose sum must satisfy the Ward--Takahashi identity.
  Not all diagrams are present for all combinations of $G$}
 \end{figure}

  Next, we consider the Ward--Takahashi identity in the set of 
amplitudes that are described by 
fig.~\ref{fig_triple_boson_current_amplitude}.
  This allows us to work out the Goldstone-boson quartic couplings:
 \begin{equation}
  \left\{\begin{array}l
   g_{G^+G^-G^0G^0}=M_H^2/4f^2,\\
   g_{G^+G^+G^-G^-}=2M_H^2/4f^2,\\
   g_{G^0G^0G^0G^0}=3M_H^2/4f^2.
  \end{array}\right.
 \end{equation}

 \begin{figure}[ht]
  \includegraphics{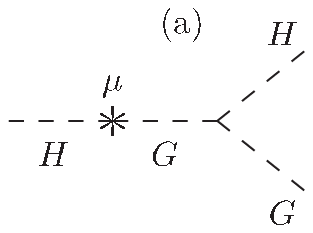}
  \includegraphics{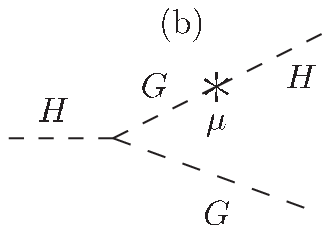}

  \includegraphics{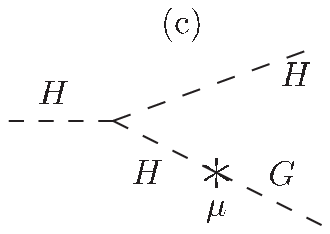}
  \includegraphics{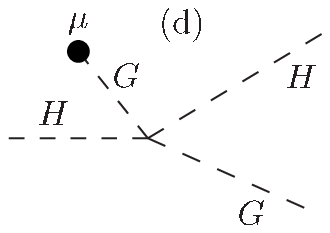}
  \caption{\label{fig_triple_boson_current_amplitude_second}
  Four diagrams whose sum must satisfy the Ward--Takahashi identity.}
 \end{figure}

  We then turn to the set of amplitudes that are described by 
fig.~\ref{fig_triple_boson_current_amplitude_second}, and obtain
 \begin{equation}
  g_{GGHH}=\frac1{f}\left(-g_{GGH}+\frac{g_{HHH}}2\right).
 \end{equation}

  Note that the Higgs-boson self-coupling is not fixed by current 
conservation conditions but either by loops or by the symmetry
between Goldstone and Higgs bosons.
  As noted in ref.~\cite{gribovewsb}, these turn out to be the same as 
the couplings of the Standard Model:
 \begin{equation}
  g_{HHH}=\frac{3M_H^2}{2f},\qquad g_{HHHH}=\frac{3M_H^2}{4f^2}.
 \end{equation}
  This implies
 \begin{equation}
  g_{GGHH}=\frac{M_H^2}{4f^2}.
 \end{equation}
  We notice that the effective Lagrangian for the multi-boson 
interaction terms can be written as
 \begin{equation}
  -\mathcal{L}_\mathrm{eff}=\frac{M_H^2}{8v^2}
  \left[
   2G^+G^-+(G^0)^2+(v+H)^2-v^2
  \right]^2.
 \end{equation}
  In terms of the Standard-Model Higgs-doublet field $\Phi$, this 
expression is proportional to $(\Phi^\dagger\Phi-v^2)^2$. That is, the 
mass-generation mechanism of the effective theory is equivalent to that 
of the Standard Model.

 \section{Tadpole cancellation}

  We now consider the tadpole cancellation condition:
 \begin{equation}
  \includegraphics{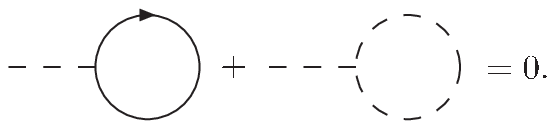}
 \end{equation}
  Let us neglect the contribution of all fermions other than the top 
quark.
  This yields
 \begin{equation}
  \int_{m_t^2}^{\lambda^2}dq^2\left[
  N_C\frac{4m_t^2(q^2)}{2f}-\frac32\frac{M_H^2(0)+M_H^2(q^2)}{2f}
  \right]=0,
  \label{eqn_tadpole_cancellation_condition}
 \end{equation}
  where the simplification $q^2\gg m_t^2, M_H^2$ has been made.
  The shorthand notation $m_t=m_t(m_t^2)$ is used.
  Note that in the Goldstone loop, $M_H$ is not corrected 
by renormalization effects.
  This is in contrast to the Higgs-boson loop, which is suppressed 
because of the running Higgs mass.
  We have neglected the renormalization effects to $f^2$, which are 
relatively small, due to the Goldstone and Higgs boson propagators.

  Since eqn.~(\ref{eqn_tadpole_cancellation_condition}) is dominated by 
the large energy region, we then obtain
 \begin{equation}
  M_H^2(0)\approx8m_t^2(\lambda^2),
  \label{eqn_tadpole_cancellation_result}
 \end{equation}
  as discussed in the introduction.
  Strictly speaking, the scale on the right-hand side should be slightly 
below $\lambda$. The running of the top quark mass is as given by 
eqn.~(10) of ref.~\cite{gribovewsb}:
 \begin{equation}
  m_t(q^2)=m_t\left(\alpha_s(q^2)/\alpha_s(m_t^2)\right)^{4/7},
 \end{equation}
  where only the QCD part of the evolution is included, and
 \begin{equation}
  \alpha_s^{-1}(q^2)=\alpha_s^{-1}(m_t^2)+\frac7{4\pi}
  \ln\frac{q^2}{m_t^2}.
 \end{equation}

 \begin{figure}[ht]
  \includegraphics[width=8.5cm]{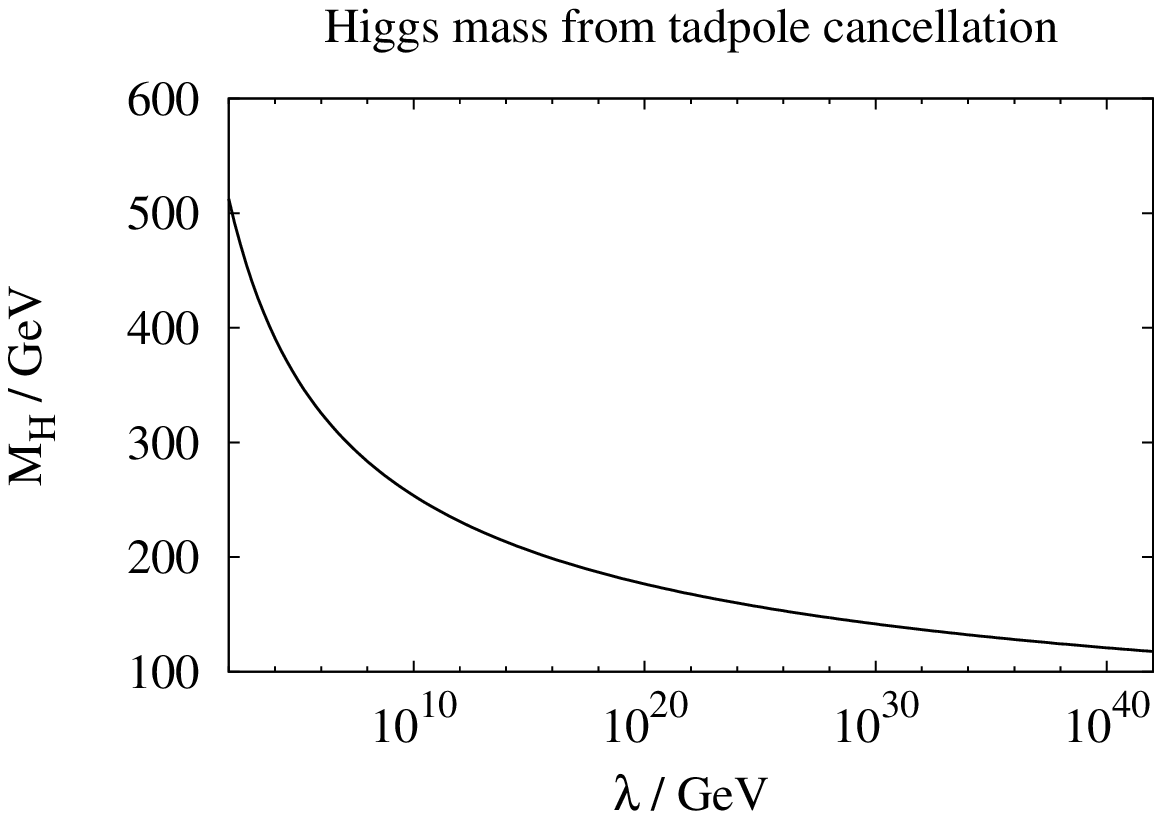}
  \caption{\label{fig_Htad}
  $M_H(0)$ as a function of the cut-off scale $\lambda$, using 
eqn.~(\ref{eqn_tadpole_cancellation_result}), for $m_t(m_t)=174$~GeV.
  Only the QCD part of the evolution is included.}
 \end{figure}

  The result of eqn.~(\ref{eqn_tadpole_cancellation_result}) is shown in 
fig.~\ref{fig_Htad}.
  The U(1)$_Y$ scenario yields $M_H=117$~GeV, when $\lambda$ 
($\sim10^{42}$~GeV) is given by eqns.~(45) and (46) of 
ref.~\cite{gribovewsb}:
 \begin{equation}
  (\alpha^\prime(\lambda^2))^{-1}=(\alpha^\prime(m_t^2))^{-1}-
  \frac{5}{3\pi}\ln\frac{\lambda^2}{m_t^2}\approx0.
 \end{equation}
  As mentioned in the introduction, this spectacular agreement with the 
LHC results \cite{cernpressrelease,atlashiggs,cmshiggs} should be 
treated with caution and suspicion.
  The full running $m_t$, which includes both QCD and U(1)$_Y$ running, 
formally vanishes at the Landau pole. The reason for our omitting the 
latter running is that the effect is much less than the QCD 
running, except at very near the Landau pole. Furthermore, at near the 
Landau pole, the perturbative result for the running mass cannot be 
trusted.

  Having said that, we may, as a means of error estimation, include the 
U(1)$_Y$ running effect in 
eqn.~(\ref{eqn_tadpole_cancellation_condition}) which is then evaluated 
literally. This yields $88$~GeV. 
Thus there is roughly 25~\% error in our prediction of $M_H=117$~GeV.

  Even so, this result provides some support to Gribov's U(1)$_Y$ 
scenario since, as can be seen from fig.~\ref{fig_Htad}, low $M_H$ 
requires large $\lambda$.

 \section{Bosonic self-energies}

\begin{figure}[ht]
  \includegraphics{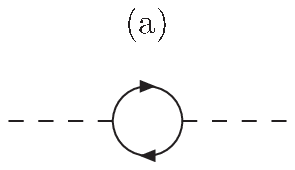}
  \includegraphics{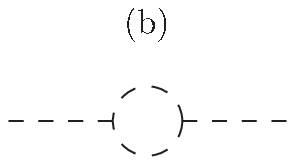}

  \includegraphics{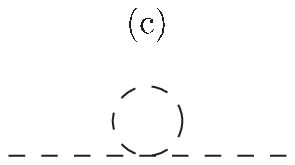}
  \caption{\label{fig_higgs_self_energy}
  The diagrams for the self-energy of the Goldstone and Higgs bosons. 
Diagram b involves the $g_{GGH}$ and $g_{HHH}$ couplings, and
diagram c involves $g_{GGHH}$, $g_{GGGG}$ and $g_{HHHH}$ couplings.}
 \end{figure}

  The three types of Feynman diagram for the self-energy of Goldstone 
and Higgs bosons are shown in fig.~\ref{fig_higgs_self_energy}.
  We have not shown the tadpole diagrams of the form discussed in the 
previous section, as these are zero if the tadpole cancellation 
condition is imposed.
  The contributions of figs.~\ref{fig_higgs_self_energy}a and c are 
quadratically divergent, and we shall show that they mutually cancel.
  The contribution of fig.~\ref{fig_higgs_self_energy}b provides a 
correction to eqn.~(\ref{eqn_gribov_higgs_mass}).

  Let us consider the Goldstone-boson self-energy.
  At zero external momentum, the contributions of the three diagrams are 
given by:
 \begin{eqnarray}
  \Sigma^G_a&=&\int\frac{d^4k}{(2\pi)^4i}
  \frac{m_t^2N_C}{4f^2}\mathrm{Tr}\left[\frac1{k^2-m_t^2}\right],\\
  \Sigma^G_b&=&-\int\frac{d^4k}{(2\pi)^4i}
  \frac{M_H^4}{4f^2}\frac1{k^2(k^2-M_H^2)},\\
  \Sigma^G_c&=&-\int\frac{d^4k}{(2\pi)^4i}
  \frac{M_H^2}{4f^2}\left[\frac5{2k^2}+\frac1{2(k^2-M_H^2)}\right].
 \end{eqnarray}
  We see that adding together these three equations yields zero so long 
as the tadpoles cancel.
  For a more formal, all-order treatment of such cancellations in the 
Goldstone-boson mass, we may, for instance, consider the non-vanishing 
part of $p^\mu\Delta\Pi_{\mu\nu}$ in eqn.~(29) of ref.~\cite{gribovewsb} 
and show that this is equivalent to tadpole contributions.
  As a generalization, we note that the divergences cancel in any case 
in Goldstone-boson self-energy, regardless of the vanishing of the 
tadpoles, if the tadpole diagrams are added explicitly to 
fig.~\ref{fig_higgs_self_energy}.

  Because the Goldstone-boson self-energy must be equal to $-p^2$, 
calculating $\Sigma^G$ at finite and small external momentum allows us 
to work out $f$, and this reproduces eqn.~(14) of 
ref.~\cite{gribovewsb}:
 \begin{equation}
  f^2=\frac{3}{2g_s^2}m_t^2\left[1-\left(
   \frac{\alpha_s(\lambda^2)}{\alpha_s(m_t^2)}
  \right)^{1/7}\right].
  \label{eqn_gribov_w_mass}
 \end{equation}
  The bosonic contributions do not give logarithmic corrections to this 
equation.
  This equation predicts $m_t$ as a function of $M_W$, or vice versa 
and, as a generic problem in top-quark condensation approaches, it is 
well known that the predicted $m_t$ is too high. For example, Gribov 
\cite{gribovewsb} predicts $m_t=215$~GeV for $\alpha_s(m_t^2)=0.11$.
  In ref.~\cite{gribovewsb}, this discrepancy is attributed to UV 
contributions near the Landau pole due to the strong dynamics.
  This is one possibility, but another source of the discrepancy may be 
new contributions (e.g., gravitational) at high scales, since 
eqn.~(\ref{eqn_gribov_w_mass}) is relatively sensitive to the mass 
evolution at high scales, unlike eqn.~(\ref{eqn_gribov_higgs_mass}) 
which is affected by $m_t$ quartically and is therefore less sensitive 
to the UV region.

  Let us now turn to the Higgs-boson self-energy. These are also given 
by the three diagrams shown in fig.~\ref{fig_higgs_self_energy}.
  The amplitudes at zero external momentum are now given by
 \begin{eqnarray}
  \Sigma^H_a&=&\int\frac{d^4k}{(2\pi)^4i}
  \frac{m_t^2N_C}{4f^2}\mathrm{Tr}\left[
  \frac{k^2+m_t^2}{(k^2-m_t^2)^2}\right],\\
  \Sigma^H_b&=&-\int\frac{d^4k}{(2\pi)^4i}
  \frac{M_H^4}{4f^2}\left[\frac3{2k^4}+\frac9{2(k^2-M_H^2)^2}\right],\\
  \Sigma^H_c&=&-\int\frac{d^4k}{(2\pi)^4i}
  \frac{M_H^2}{4f^2}\left[\frac3{2k^2}+\frac{3}{2(k^2-M_H^2)}\right].
 \end{eqnarray}
  The sum of the quadratic divergences of $\Sigma^H_a$ and $\Sigma^H_c$ 
is proportional to the tadpole and therefore vanishes. We then obtain
 \begin{equation}
  M_H^2=\frac{6}{16\pi^2f^2}
   \int_{m_t^2}^{\lambda^2}\frac{dq^2}{q^2}\left[
   m_t^4(q^2)-\frac{M_H^4(0)+3M_H^4(q^2)}{16}
  \right].
  \label{eqn_higgs_mass_modified}
 \end{equation}
  The second term of eqn.~(\ref{eqn_higgs_mass_modified}) gives a 
correction to eqn.~(\ref{eqn_gribov_higgs_mass}).

  In order to solve this integral, we need the running $M_H^2$.
  This is obtained by writing the left-hand side of 
eqn.~(\ref{eqn_higgs_mass_modified}) as $M_H^2(Q^2)$, changing the lower 
limit of integration to $Q^2$, and replacing $M_H(0)$ by $M_H(Q^2)$.
  The low-energy Higgs mass, i.e.\ $M_H(0)$, is then obtained by solving 
the resulting integral equation numerically.

 \begin{figure}[ht]
  \includegraphics[width=8.5cm]{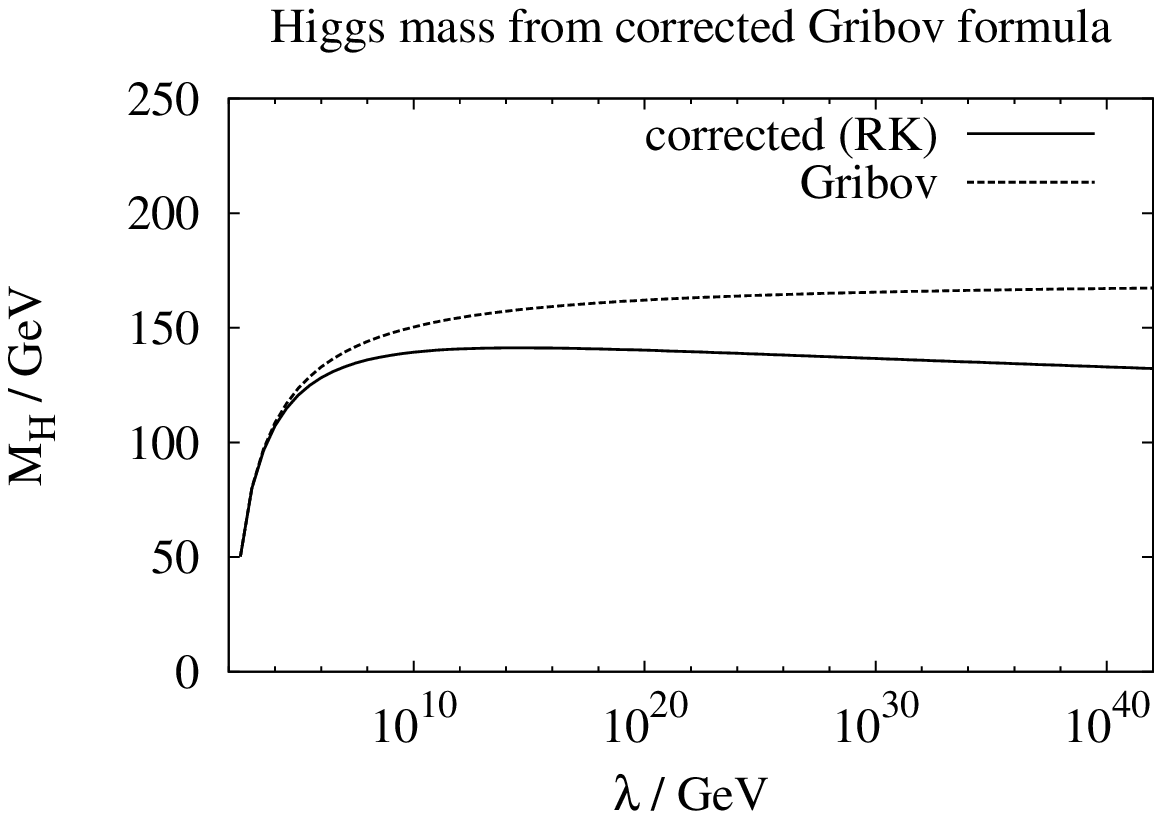}
  \caption{\label{fig_Hfull}
  $M_H$ against the cut-off scale, with the bosonic correction 
(continuous line) and without (dotted line).
  Only the QCD part of the evolution is included.}
 \end{figure}

  The result of this calculation is shown in fig.~\ref{fig_Hfull}, using 
$m_t=174$~GeV and $\alpha_S=0.11$. We see that in the U(1)$_Y$ scenario, 
$M_H$ is reduced from 167~GeV to 132~GeV.
  $M_H$ also becomes more independent of $\lambda$, and we find that
the maximum value $M_H\approx141$~GeV is obtained for
$\lambda\approx10^{15}$~GeV.

  Above $\lambda\approx10^{10}$~GeV, the uncertainty, which may be 
estimated by varying $m_t$ and $\alpha_S$, becomes nearly constant. For 
example, by varying $m_t$ by $\pm5$~GeV, we obtain $\pm5$~GeV variation 
in $M_H$, and by varying $\alpha_S$ by $\pm0.01$, we again obtain 
$\pm5$~GeV variation in $M_H$. The combined error is thus $\pm7$~GeV.
  Other sources of uncertainty are: the higher-order corrections which 
are expected to be small when the running couplings are small; the UV 
correction near the Landau pole; and other UV interactions which may be 
present.

  Note that the Higgs mass prediction of fig.~\ref{fig_Htad} is 
independent of the prediction of fig.~\ref{fig_Hfull}.
  The former was obtained by the condition that the tadpole vanishes, 
and the latter was obtained by evaluating the self-energy integrals.
  If both predictions are literally true, it follows that the point at 
which the two predictions match, which comes out to be about 
$10^{32}$~GeV, is the true scale of the UV cut-off.

\begin{figure}[ht]
  \includegraphics[width=8.5cm]{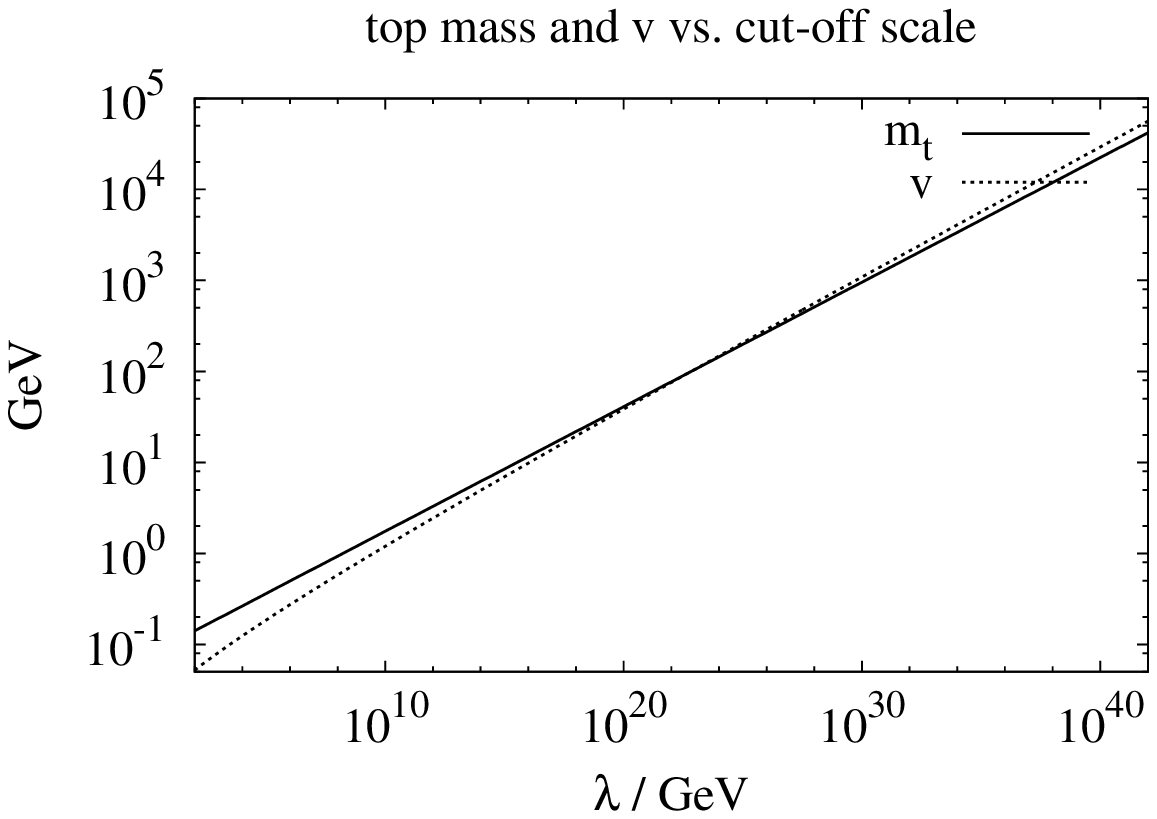}
  \caption{\label{fig_tvsl}
  $m_t$ and $v$ as a function of the cut-off scale $\lambda$, using 
eqn.~(\ref{eqn_hierarchy_generation}) and 
eqn.~(\ref{eqn_gribov_w_mass}).
  Only the QCD part of the evolution is included.}
 \end{figure}

  Another way of saying the same thing is that given a value of the UV 
cut-off, and the running coupling constants, we can work out all three 
of $M_W$, $m_t$ and $M_H$ as opposed to the previous studies in which 
$M_H$ is predicted for given $m_t$ and the cut-off, and so on.
  This follows because we now have three equations, namely eqns.~ 
(\ref{eqn_tadpole_cancellation_condition}), (\ref{eqn_gribov_w_mass}) 
and (\ref{eqn_higgs_mass_modified}), as opposed to two.
  The combined set of equations may be solved by solving the following 
integral equation:
 \begin{equation}
  \mu(r)=\int_r^R
  \frac{(R/r^\prime)^{16/7}-4(\mu^2(r)+3\mu^2(r^\prime))}
  {14R(R^{1/7}-1)}dr^\prime,
  \label{eqn_hierarchy_generation}
 \end{equation}
  numerically, with the condition $\mu(R)=0$. We then search for a value 
of $R$ that satisfies $\mu(1)\equiv M_H^2/8m_t^2(\lambda^2)=1$.
  By doing so, we obtain
 \begin{equation}
  R=\alpha_s(m_t^2)/\alpha_s(\lambda^2)=7.31.
 \end{equation}
  This ratio corresponds to a significantly large hierarchy.
  We plot the resulting $m_t$ and $v$ as a function of the cut-off scale 
$\lambda$, in fig.~\ref{fig_tvsl}.
  $M_H$ is indistinguishable from $m_t$ on the logarithmic scale.

  Taking $\lambda$ to be the U(1)$_Y$ Landau scale yields
 \begin{equation}
  m_t=4.29\times10^{4}\ \mathrm{GeV},
 \end{equation}
  which is only two orders of magnitude away from the top-quark mass 
scale.
  In order to cover the two remaining orders of magnitude, we will need 
to resolve the $m_t/M_W$ discrepancy discussed below 
eqn.~(\ref{eqn_gribov_w_mass}).
  We should mention here that the variations of $m_t$ and $\alpha_S$ 
cannot explain the two-orders-of-magnitude difference.

  One should beware of hasty conclusions, from fig.~\ref{fig_tvsl}, that 
the cut-off scale is at $\sim10^{24}$~GeV.
  The numbers that are shown are sensitive to possible corrections, 
e.g., the choice of energy scale, in 
eqn.~(\ref{eqn_tadpole_cancellation_condition}), and the need to produce 
a value of $v/m_t$ that is as large as the observed value favours large 
$\lambda$.

 \section{Discussions}

  Our result shows that the electroweak scale is not necessarily fixed 
by the parameters at the UV scale. Rather, it is fixed by dimensional 
transmutation, and this in turn is governed by the condition of vacuum 
stability and the radiative effects.

  What conditions are necessary, then, in order that symmetry breaking 
occurs in the first place?

  In our opinion, it is sufficient that the symmetry is broken by a 
chiral condensate which appears at a high scale.
  This requires that a coupling constant becomes large.
  Gribov has shown \cite{gribovquarkconfinement} that supercriticality 
and chiral symmetry breaking occurs when the relevant coupling constant 
satisfies
 \begin{equation}
  \alpha/\pi\gtrsim0.137.
 \end{equation}
  Gribov's U(1)$_Y$ scenario is a natural candidate, but one will then 
need to show that the gravitational coupling will not satisfy this 
condition.
  A possibility would be that gravitation is described by a 
weak-coupling theory above the Planck scale, becomes
strongly coupled near the Planck scale, and 
reduces to Einstein gravity at low scales.
  For example, it may be that scale generation at the U(1)$_Y$ pole
is itself the source of scale generation in gravity.

  It is interesting to ask, what might be the behaviour of the U(1)$_Y$ 
coupling at high scales? This question was addressed partially in 
ref.~\cite{odagirigribov}. We argued that when the U(1)$_\mathrm{EM}$ 
coupling grows large, it will decouple as $\alpha\propto q^2$. The 
effective coupling experienced by the electrons is then given by
 \begin{equation}
  \left(1-\frac{d\ln\alpha}{d\ln q^2}\right)\alpha\longrightarrow
  \frac1{\left|b_0\right|},
 \end{equation}
  where $b_0$ is the first coefficient of the beta function. If we can 
use the U(1)$_Y$ beta function here, then the effective coupling will 
tend to $3\pi/5$, which is large compared with $0.137\pi$. Thus chiral 
symmetry breaking will almost certainly occur by U(1)$_Y$.

  The condensate which appears at high scales must have decayed to 
fermions and Goldstone bosons at low scales, from phenomenological 
reasons. First, the observed masses of particles are light. Second, a 
condensate will lead to a cosmological constant which is much heavier 
than is observed.

  If EWSB is due to the formation of the supercritical condensate at 
high scales, one must ask the question of whether the supercritical 
states might affect the running of the parameters. Our answer is that 
the Goldstone and the Higgs states are supercritical states, but the 
other states cannot affect the running, because the Goldstone and the 
Higgs are the only states whose masses are protected by symmetry from 
growing large. The other supercritical states will have masses that are 
of the order of $\lambda$.

 \section{Conclusions}

  We have shown that Gribov's cancellation condition in top-quark 
condensation may be rephrased as the cancellation, between the top, 
Higgs and the Goldstone contributions, of the Higgs-boson tadpole.
  This is a physical condition which must be satisfied in order that the 
ground state is stable.
  The low-energy phenomenology following from our analysis is equivalent 
to that of the Standard Model.

  Tadpole cancellation gives us an extra condition for the mass of the 
Higgs boson.
  In Gribov's U(1)$_Y$ scenario, we obtain $M_H\approx117$~GeV. This is 
in good agreement with the recent LHC announcement, $M_H\approx125$~GeV, 
but our numbers are subject to $\sim 25$~\% error due to the running of 
the top-quark mass in the UV region.
  We have, furthermore, been able to calculate the bosonic contribution 
to Gribov's mass formula, eqn.~(\ref{eqn_gribov_higgs_mass}), which 
reduces Gribov's prediction of $M_H=167$~GeV to $132$~GeV.
  The error here is estimated to be $\pm7$~GeV.

  The two predictions of $M_H$ are independent.
  We may, by requiring that the two predictions match, work out the 
cut-off scale, which comes out to be $\sim10^{32}$~GeV.
  Alternatively, we may work out the low-energy parameters for a given 
value of the cut-off scale. In Gribov's scenario, we obtain 
$m_t\sim40$~TeV, which is remarkably close to the actual scale, starting 
from $\lambda\approx10^{42}$~GeV, with no other input than the values of 
the dimensionless couplings and their running.

 \section*{Acknowledgement}

  K.~O.\ thanks R.~Sinha and IMSc Chennai, where this work was carried 
out, for their warm hospitality.

  We benefited from discussions with I.~Hase, T.~Mondal, R.~C.~Verma, 
K.~Yamaji and T.~Yanagisawa.

 \end{document}